\documentclass{article}
\usepackage{spconf,amsmath,graphicx,cite,bm,booktabs,bbding,textcomp,amssymb,url}
\let\oldbibliography\thebibliography
\renewcommand{\thebibliography}[1]{%
  \oldbibliography{#1}%
  \setlength{\itemsep}{0pt}%
}

\title{Dynamic curriculum learning via data parameters \\for noise robust keyword spotting}
\name{Takuya Higuchi, Shreyas Saxena, Mehrez Souden, Tien Dung Tran, Masood Delfarah and Chandra Dhir}
\address{Apple}

\begin{document}
%
\maketitle
\begin{abstract}
 We propose dynamic curriculum learning via data parameters for noise robust keyword spotting. Data parameter learning has recently been introduced for image processing, where weight parameters, so-called data parameters, for target classes and instances are introduced and optimized along with model parameters. The data parameters scale logits and control importance over classes and instances during training, which enables automatic curriculum learning without additional annotations for training data. Similarly, in this paper, we propose using this curriculum learning approach  for acoustic modeling, and train an acoustic model on clean and noisy utterances with the data parameters. The proposed approach automatically learns the difficulty of the classes and instances, e.g. due to low speech to noise ratio (SNR), in the gradient descent optimization and performs curriculum learning. This curriculum learning leads to overall improvement of the accuracy of the acoustic model. We evaluate the effectiveness of the proposed approach on a keyword spotting task. Experimental results show 7.7\% relative  reduction in false reject ratio with the data parameters compared to a baseline model which is simply trained on the multiconditioned dataset.
\end{abstract}

\vspace{-6pt}
\section{Introduction}
\vspace{-7pt}
Acoustic modeling is essential for speech applications, such as keyword spotting and automatic speech recognition, e.g. \cite{chen2014small, sainath2015convolutional, team2017hey, he2017streaming, hinton2012deep,sainath2013deep,abdel2014convolutional, chiu2018state}. Noise robustness of an acoustic model is one of critical points for successful acoustic modeling particularly in far field speech scenarios \cite{kinoshita2013reverb,barker2015third,barker2018fifth,haeb2019speech}.
 Various approaches have been proposed, including front-end speech enhancement \cite{heymann2016neural,erdogan2016improved,yoshioka2015ntt,higuchi2016robust,boeddeker2018front} and data augmentation for acoustic modeling \cite{seltzer2013investigation,qian2016very,cui2015data}. One promising approach for improving the noise robustness is data augmentation using noise signals, where the noise signals are added to clean utterances in training data. This data augmentation, also called multicondition training, artificially introduces difficult training samples, which typically leads the acoustic model to be more robust against noise. However, it is unclear what degree of difficulty would help to generalize the model the most, so it is difficult to understand how to obtain the maximum benefit from the multicondition training. Although the augmented training data have various degrees of difficulty for acoustic modeling, e.g. high/low SNRs, and easy/hard phone states for classification, it is difficult to design an efficient training curriculum exploiting the variability of the augmented training data.

Recently, data parameters were proposed for curriculum learning, and their effectiveness was demonstrated on image classification  \cite{saxena2019data}. Weight parameters are introduced for target classes and instances, i.e. image samples, in training data. The data parameters are used to scale logits depending on target classes and instances, and are optimized with the model parameters. By scaling the logits, we can control the contribution of each class/instance to the gradients. Optimizing the data parameters during training enables class-level and/or instance-level dynamic curriculum learning by dynamically updating the scaling factors. 

The main contributions of this paper are: 1) an application of the data parameter to acoustic modeling for keyword spotting, and 2) a combination of the data parameter approach with a data augmentation technique to yield further performance improvement. In an acoustic modeling scenario, class parameters are introduced for the target phoneme classes, which vary depending on time frames and control contributions of the target classes. On the other hand, instance parameters are introduced depending on utterances, which control contributions of the utterances based on their hardness, e.g. clean/noisy.
In our experimental evaluation on a keyword spotting task, the data parameter approach combined with data augmentation yielded up to  7.7\% relative improvement in terms of the false reject ratio (FRR) compared to a baseline model trained on a noisy data.



\vspace{-6pt}
\section{Data parameters for acoustic modeling}
\label{sec:data_param}

\vspace{-8pt}
\subsection{Class parameters}
\vspace{-4pt}
Let $\lbrace \bm{x}_t^{i}, y_t^{i}\rbrace$ denote the data, where $\bm{x}_t^{i}$ denotes an audio feature vector at time $t \in\lbrace 1, ..., T^{i}\rbrace$ of utterance $i \in\lbrace 1, ..., I\rbrace$, and $y_t^{i} \in \lbrace 1, ..., K\rbrace$ denotes its corresponding class label, e.g. a phone state label. During training, an acoustic model, i.e. a deep neural network (DNN) takes the input $ \bm{x}_t^{i}$ and estimates logits $\bm{z}_t^{i}$ as $\bm{z}_t^{i} = f_{\theta}(\bm{x}_t^{i})$, where $\theta$ denotes model parameters and $f_{\theta}$ denotes a mapping function by the DNN.

Instead of computing the softmax directly using the logits, we introduce the class parameters to scale the logits. Let $\sigma^{class} = \lbrace\sigma_{1}^{class}, ..., \sigma_{K}^{class}\rbrace$ denote the class parameters for the target classes $1, ..., K$. The probability for target class $y_t^{i}$ at time $t$ of utterance $i$ can be written as
\begin{equation}
p_{t, y_t^{i}}^{i} = \frac{exp(z_{t, y_t^{i}}^{i}/\sigma_{y_t^{i}}^{class})}{\sum_{j}exp(z_{t, j}^{i}/\sigma_{y_t^{i}}^{class})},\label{pb_class}
\end{equation}
where $z_{t, y_t^{i}}^{i}$ and $\sigma_{y_t^{i}}^{class}$ denote logits and the class parameter for target class $y_t^{i}$, respectively. Note that the class parameter used in both the numerator and denominator is determined by target class $y_t^{i}$. Therefore, the class parameter cannot be absorbed into model parameters in contrast to approaches introducing scaling parameters into model parameters (e.g.  \cite{ghahremani2016linearly}). Thus, the class parameters control contributions of the target classes during training, keeping the model architecture unchanged.  When we set all the class parameters to one, i.e. $\sigma_{j}^{class} = 1, j = 1, ..., K$, Eq. (\ref{pb_class}) is equivalent to the standard softmax computation.

\vspace{-8pt}
\subsection{Instance parameters}
\vspace{-4pt}
Instance parameters are specific to a particular utterance, so they control curriculum over instances. Let $\sigma^{inst} = \lbrace\sigma_{1}^{inst}, ..., \sigma_{I}^{inst}\rbrace$ denote the instance parameters for utterances $1, ..., I$ in training data. Using the instance parameters, the scaled probability can be written as:
\begin{equation}
p_{t, y_t^{i}}^{i} = \frac{exp(z_{t, y_t^{i}}^{i}/\sigma_{i}^{inst})}{\sum_{j}exp(z_{t, j}^{i}/\sigma_{i}^{inst})}.\label{pb_inst}
\end{equation}
Note that the same instance parameter is used across time frames in the same utterance, which allows us to perform utterance-wise curriculum learning. 

\vspace{-8pt}
\subsection{Data parameters}
\vspace{-4pt}
The data parameters are defined as the sum of the class and  instance parameters, $\sigma_{t,i}^{*} = \sigma_{y_t^{i}}^{class} + \sigma_{i}^{inst}$. The data prameters are used to scale the logits similarly to Eqs. (\ref{pb_class}) and (\ref{pb_inst}). 
Unlike the data parameters for image processing \cite{saxena2019data}, the data parameters for speech processing have different values at different time frames depending on both the class and instance parameters, even if data points at the time frames are from the same utterance. By optimizing both the class and instance parameters, we can perform curriculum learning over both classes and instances.

\vspace{-8pt}
\subsection{Data parameter optimization}
\vspace{-4pt}
The data parameters can be optimized via a gradient descent algorithm, and we use the usual averaged cross entropy loss over the time frames and the instances:
\begin{align}
L &= - \frac{1}{T^{*}} \sum_{t,i} L_{t}^{i} \\ 
 &= - \frac{1}{T^{*}} \sum_{t,i} \log(p_{t, y_t^{i}}^{i}), \label{obj}
\end{align}
where $T^{*}$ denotes the total number of time frames in training data, and $p_{t, y_t^{i}}^{i}$ is the probability scaled by the data parameter, $\sigma_{t,i}^{*}$. We minimize $L$ with respect to model parameters, $\theta$, as well as the class and instance parameters. The gradient of the loss with respect to logits can be written as:
\begin{equation}
\frac{\partial L_{t}^{i}}{\partial z_{t,j}^{i}} = \frac{p_{t,j}^{i} - 1(j=y_t^{i})}{\sigma_{t,i}^{*}}. \label{partial_logits}
\end{equation}
Thus the gradient gets scaled by $\sigma_{t,i}^{*}$, where different scaling factors are applied at different data points. The data parameter values control importance over different time frames and utterances during model parameter optimization.

On the other hand, the gradient with respect to the data parameters can be written as:
\begin{equation}
\frac{\partial L_{t}^{i}}{\partial \sigma_{t,i}^{*}} = \frac{(1 - p_{t,y_t^{i}}^{i})}{(\sigma_{t,i}^{*})^2}(z_{t, y_t^{i}}^{i} - \sum_{j\neq y_t^{i}}q_{t,j}^{i}z_{t, j}^{i}), \label{partial_data}
\end{equation}
where $q_{t,j}^{i} = \frac{p_{t,j}^{i}}{1-p_{t,y_t^{i}}^{i}}$ is the probability distribution over non-target class $j$ scaled by $1-p_{t,y_t^{i}}^{i}$. Since $\sum_{j\neq y_t^{i}}q_{t,j}^{i}z_{t, j}^{i}$ corresponds to the expected value on non-target classes, the data parameter $\sigma_{t,i}^{*}$ will increase if the logit of the target class $z_{t, y_t^{i}}^{i}$ is smaller than the expected logits on non-target classes and vice-versa. The gradients for the class and instance parameters can also be written as Eq. (\ref{partial_data}) since the data parameter is formed by the addition of these parameters.

From Eqs. (\ref{partial_logits}) and (\ref{partial_data}), if the DNN misclassifies a data point, the data parameter will gradually increase, which will decay the gradient of the logit. Decreasing the data parameter has an opposite effect, and  accelerate the optimization of the model parameters. This mechanism helps the DNN to focus on easy data points at the beginning of training, and automatically leaves the harder cases until later based on the performance of the model.

We optimize $\log(\sigma^{class})$ and $\log(\sigma^{inst})$ instead of $\sigma^{class}$ and $\sigma^{inst}$ to avoid them having negative values. In addition, we have $l_{2}$ regularization on the data parameters, i.e. $\parallel \log(\sigma^{class} + \sigma^{inst}) \parallel ^{2}$ to prevent the data parameters from having too high or low values, and to favor the original softmax computation with $\sigma^{*} = 1$. The contribution of this regularization in the objective function can be controlled by a weight decay hyperparameter.

\vspace{-6pt}
\section{Experimental evaluation}
\label{sec:exp}

\vspace{-8pt}
\subsection{Data}
\vspace{-4pt}
Clean training data consists of $\sim$500k utterances in English recorded anonymously with a smart speaker. The utterances contains a trigger phrase, i.e. ``Hey Siri" followed by a query. The trigger phrase and the query parts correspond to positive and negative samples, respectively, for model training. To create noisy training data, we used household noises and noise samples taken from the BBC sound effect dataset \cite{BBC}, including vacuum, grinders, screwdriver, hammer, workshop, alarms, etc. These noise signals were convolved with room impulse responses that were measured in typical residential environments with varying reverberation times and locations of the sources and microphone array. 
  We added the reverberated noise signals to the clean utterances to obtain a certain SNR, which was randomly sampled from a uniform distribution ranging [-10dB, 10dB). By appending the noisy utterances to the clean utterances, we created noisy training data consisting of $\sim$1000k utterances. We used $\sim$2\% of the data for cross validation to tune hyperparameters.

Evaluation data consisted of 13186 positive utterances which have the trigger phrase, and 2000 hours of negative data obtained by playing TV, radio, and podcast in rooms. All the evaluation data were recorded with the same microphone array. The positive utterances were recorded in realistic room conditions, e.g. with room noises and/or playback with various volumes. Our front-end speech enhancement system \cite{Applefrontend} was applied to both training and evaluation data.

\vspace{-8pt}
\subsection{Settings}
\vspace{-4pt}

We used 5 layers of fully-connected networks with 64 units. Batch normalization was applied after each fully-connected layer, followed by the sigmoid activation function. The last layer projected the 64-dimensional hidden activation to logits for the target classes. An acoustic model with this size can be always-on for keyword spotting in streaming audio on a smart speaker.
   The target classes consisted of 18 classes (3 states $\times$ 6 tri-phones) for the trigger phrase, one for silence and non-speech noise class, and one for other speech class. During inference, we used an HMM to accumulate probabilities for the keyword phone sequence to obtain a keyword spotting score. The transition probabilities of the HMM were obtained from the distributions of training data. See \cite{sigtia2018} for more details of DNN-HMM based keyword spotting. We used 13-dimensional mel-frequency cepstral coefficients (MFCCs) for input features. The features from the previous 9 frames and the future 9 frames were concatenated to the feature vector at the current frame, which resulted in 13x19=247-dimensional feature vector per frame. Adam optimizer \cite{kingma2014adam} was used for model parameter optimization. The initial learning rate was set at $0.01$, and betas were set at [0.9,0.999]. A learning rate decay of 0.5 was applied when the cross validation loss did not decrease at two consecutive epochs. Early-stopping was applied when we did not see cross validation loss decrease at 9 consecutive epochs. We used $256$ utterances per minibatch for training. Two baseline models were trained on the clean and noisy data, respectively, without using the data parameters. Regarding the data parameters, we considered three settings: 1) class parameter only, 2) instance parameter only, and 3) joint class and instance parameters. For the data parameter optimization, stochastic gradient descent (SGD) was used with the weight decay. To avoid the class and instance parameters becoming zero or having high values during optimization, clipping was applied to the class parameters with a minimum value of $0.05$ a maximum value of $20$, and to the instance parameters with a minimum value of $0.0001$ and a maximum value of $20$. No learning rate decay was applied for the class and instance parameters.  The data parameter implementation was done based on the open source software \footnote{https://github.com/apple/ml-data-parameters} provided by the authors of \cite{saxena2019data}.

\begin{table}[t!]
\vspace{-8pt}
\setlength{\abovecaptionskip}{5pt}
  \caption{Hyperparameters for data parameters}
  \label{tab:hyperparam}
  \centering
  \scalebox{0.9}{
  \begin{tabular}{ r r |r r|r r|r }
    \toprule
    \multicolumn{1}{c}{\textbf{}} & 
    \multicolumn{1}{c}{\textbf{}} & 
                                         \multicolumn{2}{|c}{\textbf{class}} &
                                         \multicolumn{2}{|c}{\textbf{inst.}} &
                                         \multicolumn{1}{|c}{\textbf{}}\\
    \multicolumn{1}{c}{\textbf{data}} & 
    \multicolumn{1}{c}{\textbf{case}} &
                                         \multicolumn{1}{|c}{\textbf{lr}} &
                                         \multicolumn{1}{c}{\textbf{init}} &
                                         \multicolumn{1}{|c}{\textbf{lr}} &
                                         \multicolumn{1}{c}{\textbf{init}} &
                                         \multicolumn{1}{|c}{\textbf{wd}}\\
    \midrule
     & class &0.001 &1&&&0.01          \\
    clean & inst. & &&0.001&1& 0.01            \\
     & joint&0.001 &1&0.1&0.01& 0.01          \\
     \hline
          & class &0.001 &1&&&   0.01 \\
    noisy & inst. & &&0.01&1& 0.1           \\
     & joint&0.001 &1&1&0.1&     0.01       \\

    \bottomrule
  \end{tabular}
  }
  \vspace{-8pt}
\end{table}

\begin{figure}[t!]
\setlength{\abovecaptionskip}{-5pt}
  \centering
  \includegraphics[width=0.85\linewidth]{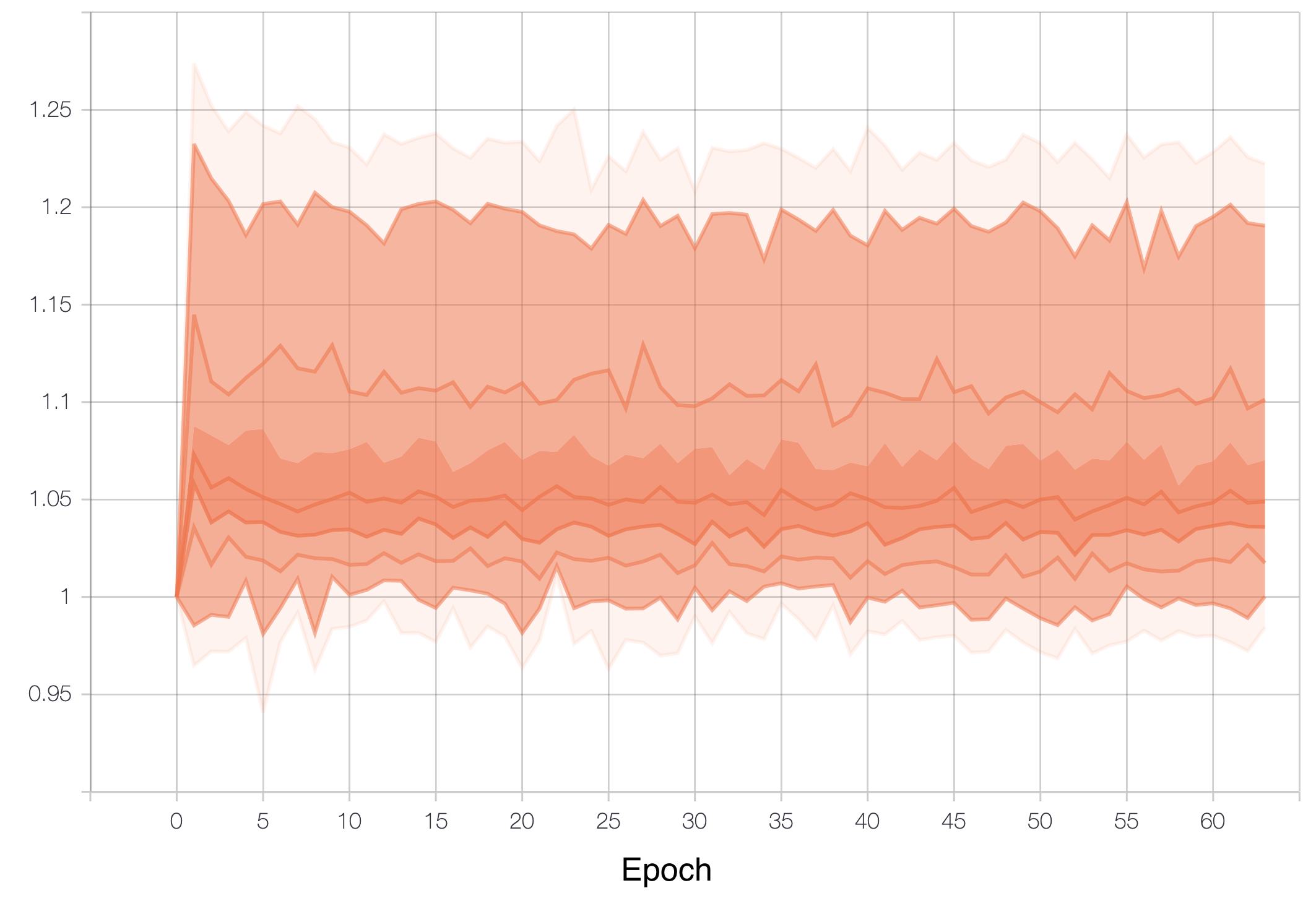}
  \vspace{-7pt}
  \caption{A distribution of the class parameters during training. The line in the middle shows the median, and the colored regions have widths $\sigma$, 2$\sigma$ and 3$\sigma$ respectively, where $\sigma$ denotes a standard deviation of the distribution. The lowest and the highest lines indicate the minimum and maximum values respectively.}
  \label{fig:class_dist}
\end{figure}

Hyperparameters for the data parameters, i.e. learning rate, initial value, and weight decay were set as shown in Table \ref{tab:hyperparam}. The weight decay parameter was multiplied by the $l_{2}$ regularization term to control its contribution to the objective function. For the joint cases, we kept the same learning rate and the same weight decay as used in the class parameters only case, and tuned the learning rate and the initial values for the instance parameters. The initial values were set so that the sum of the class and instance parameters kept close to $1$.  

All models including baseline were trained using the same random seed for model initialization and data ordering, and hyperparameter search was performed based on the cross validation loss.

\vspace{-8pt}
\subsection{Distributions of class and instance parameters}
\vspace{-4pt}

Figure \ref{fig:class_dist} shows an example of a distribution of the class parameters during training, which was obtained with the noisy data in the class parameter only case. Starting with the initial value of $1$, the class parameters spread higher/lower at the beginning part of training, and then converged into certain values depending on the target classes. This distribution shows that target class-level curriculum learning via the class parameters was performed during training. Similar distributions were observed with the other settings.

\begin{figure}[t]
\vspace{-13pt}
  \centering
  \includegraphics[width=0.9\linewidth]{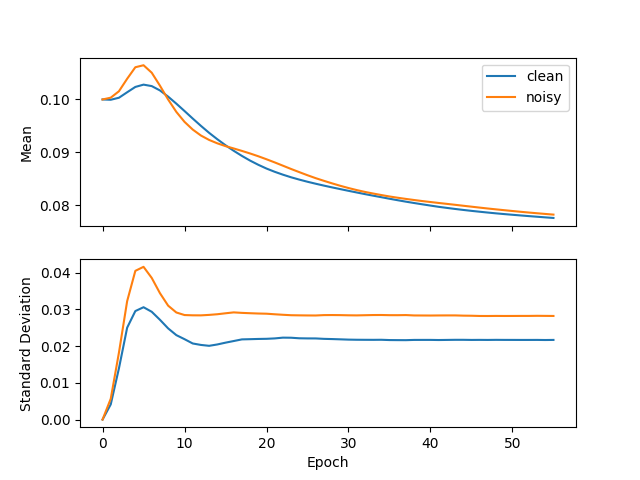}
  \vspace{-7pt}
  \caption{Means and standard deviations of distributions of the instance parameters for the clean and noisy utterances.}
  \setlength{\abovecaptionskip}{-15pt}
  \label{fig:inst_param_avg_both}
  \vspace{-8pt}
\end{figure}

Figure \ref{fig:inst_param_avg_both}  shows means and  standard deviations of distributions of the instance parameters for the clean and the noisy utterances in the noisy data, respectively. The distributions obtained in the joint case are shown. At the beginning of training, the mean for the noisy utterances increases more than the mean for the clean utterances, then both converge to similar values. This means that, on average, the model focused more on the clean utterances at the early stage of training, and then started treating equally the clean and noisy utterances. The standard deviation for the noisy utterances was higher at the beginning and converged at a higher value compared with that for the clean utterances. This broader distribution on the noisy data indicates that the instance parameters controlled the importance of the noisy utterances depending on the degrees of difficulty of the noisy utterances.  A similar tendency was observed with other settings as well. Note that clean/noisy annotations were not used during training, but were used only to calculate the instance parameter distributions separately on the clean and noisy utterances for this figure.

\vspace{-8pt}
\subsection{False reject ratios}
\vspace{-4pt}

\begin{table}[t]
\vspace{-14pt}
  \caption{False reject ratios at 10 false alarm per hour using clean training data}
  \label{tab:FRRs_clean}
  \centering
  \scalebox{0.9}{
  \begin{tabular}{ r l  r |r |r }
    \toprule
    \multicolumn{1}{c}{\textbf{Models}} & 
                                         \multicolumn{1}{c}{\textbf{class}} &
                                         \multicolumn{1}{c}{\textbf{inst.}} &
                                         \multicolumn{1}{c}{\textbf{FRRs (rel. impr.) [\%]}}\\
    \midrule
    baseline & & &2.37~~~~~~~~~~~~~             \\
    class param.                      & \checkmark &  & \bf{2.28} ~\bf{(3.8)}~~~               \\
    inst. param.                    & & \checkmark  & 2.40 (-1.3)~~~       \\
    joint              &\checkmark & \checkmark   & 2.47 (-4.2)~~~              \\
    \bottomrule
  \end{tabular}
  }
  \vspace{-8pt}
\end{table}

Table \ref{tab:FRRs_clean} shows the false reject ratios (FRRs) at a certain operating point, i.e., 10 false alarm  (FA) per hour\footnote{10 FA per hour is a reasonable operating point for always-on models on device with this size, being evaluated on this negative data. The FAs can be mitigated by using muti-stage approaches, e.g., \cite{gruenstein2017cascade,AppleHeySiri}.} obtained with the clean training data. Although the class parameters improved the performance, we could not get substantial improvements when using the instance parameters with the clean training data. The reason could be that the clean data did not have a moderate diversity of difficulties across data samples, hence the instance parameters introduced the unnecessarily high degree of freedom which could result in overfitting.


\begin{table}[t!]
\setlength{\abovecaptionskip}{5pt}
  \caption{False reject ratios at 10 false alarm per hour using noisy training data}
  \label{tab:FRRs_noisy}
  \centering
  \scalebox{0.9}{
  \begin{tabular}{ r l  r |r |r }
    \toprule
    \multicolumn{1}{c}{\textbf{Models}} & 
                                         \multicolumn{1}{c}{\textbf{class}} &
                                         \multicolumn{1}{c}{\textbf{inst.}} &
                                         \multicolumn{1}{c}{\textbf{FRRs (rel. impr.) [\%]}}\\
    \midrule
    baseline & & &2.22~~~~~~~~~~~~             \\
    class param.                      & \checkmark &  & ~~~2.06 (7.2)~~~               \\
    inst. param.                    & & \checkmark  & 2.13 (4.1)~~~       \\
    joint              &\checkmark & \checkmark   & \bf{2.05} \bf{(7.7)}~~~              \\
    \bottomrule
  \end{tabular}
  }
  \vspace{-8pt}
\end{table}

Table \ref{tab:FRRs_noisy} shows the FRRs obtained using the noisy training data. By simply training the model on the noisy data, the FRR improved by 6.3\% relative to the baseline model trained on the clean data. By using the joint class and instance parameters, we achieved further 7.7\% improvement relative to the model trained on the noisy data without using the data parameters, although the improvement by the joint optimization was limited compared to the result with only the class parameters.  In contrast to the results with the clean training data, the instance parameters also improved the FRRs from the baseline by 4.1\% relative. These results shows the effectiveness of the proposed approach on the keyword spotting task in a multicondition training scenario.

\vspace{-6pt}
\section{Related work}
\label{sec:related}
\vspace{-7pt}
Sivasankaran et al. \cite{sivasankaran2017discriminative} also proposed utterance-wise weight parameters for multicondition training, where they directly scaled the cross entropy loss using the weight parameters.
 However, in their experiments, they split training data based on SNR, and used subset-wise weight parameters noting that utterance-wise weights are undesirable and would result in overfitting. In this work, we introduced utterance-wise data parameters and demonstrated how they improved the accuracy without additional annotations, such as SNR labels.

\vspace{-6pt}
\section{Conclusions}
\label{sec:conc}
\vspace{-7pt}
We proposed data parameters for noise robust keyword spotting. The data parameters consisted of the class and instance parameters, which controlled the contributions of the target classes and the instances during training.
 Experimental results with  augmented training data showed that the data parameter approach achieved up to 7.7\% relative improvement over the baseline model simply trained on the noisy data.


\bibliographystyle{IEEEbib_short}
\bibliography{mybib_short_et_al}

\end{document}